\pdfoutput=1  % arXiv: force pdfLaTeX (ebgaramond + other font packages need it, not the dvi route)
\documentclass[10pt,twocolumn]{article}
\usepackage[utf8]{inputenc}
\usepackage[T1]{fontenc}
\usepackage[margin=0.75in,columnsep=0.3in]{geometry}
\usepackage{ebgaramond}
\usepackage{ebgaramond-maths}
\usepackage{booktabs}
\usepackage{tikz}
\usepackage[font=small,labelfont=bf,labelsep=period]{caption}
\usepackage{microtype}
\usepackage{enumitem}
\usepackage[hidelinks]{hyperref}
\usepackage{listings}
\newcommand{\tayyardoi}{10.5281/zenodo.23704154}% the Zenodo DOI of the v1.0 release
\newcommand{\doinote}{\ifx\tayyardoi\empty\ and its archival DOI listed on the same page\else\ and archived under DOI \href{https://doi.org/\tayyardoi}{\tayyardoi}\fi}

\title{\textbf{The Model as One Rater Among Several:}\\
Measuring Political Positions in Data-Sparse Regions\\ with a Language-Model Panel}
\author{Tarek Gara}
\date{Version 2, October 2026}

\begin{document}
\twocolumn[{%
\maketitle
\begin{abstract}
\noindent Most instruments for measuring political positions, such as manifesto coding, expert surveys and text-scaling models, were built and validated on Western party systems, and outside that setting they work poorly or not at all. They break down worst in regions where many consequential actors never publish a manifesto, never contest a competitive election, or commit one position to a charter and act on another in office. This paper proposes a method for such settings. I treat a large language model as a single fallible rater in a panel, the way an expert survey treats one expert, so the value of the measurement comes from pooling many judges. I describe the panel and how its scores are combined, an applicability rule that keeps a blank distinct from a score of zero, and a lens system that separates what an actor declares from what it does, and I report three results. First, a built-in experiment isolates the effect of the rubric. With a definition-free first round held fixed, adding written axis definitions moves scores by a mean of 1.8 points on a 21-point scale and tightens agreement between two raters on the same cells (mean absolute gap $2.81 \rightarrow 2.50$; $r\ 0.81 \rightarrow 0.89$). An arbitrary steer would move the scores without bringing independent raters closer, so the definitions are clarifying the scale. Second, across nine models from eight laboratories in two countries, agreement is high on an interval and on an ordinal metric (Krippendorff's $\alpha$ = 0.86 and 0.87), and it held as the panel grew from five raters to nine. This is reliability, the reproducibility of a reading, and it says nothing yet about validity, since models trained on overlapping data can share a bias that no statistic computed from the models alone can detect. Third, the panel's disagreement carries information. Its sharpest split, a bimodal divergence across the whole scale on an actor's stance toward the state's foundational order, comes from what I call a referent problem. For a party at a moment of constitutional rupture, the models agree on the facts and differ on which order counts as the regime, the one the actor toppled or the one it now defends. A blind triple-coding of the widest such cells traces about two-thirds of the disagreement to interpretation, and no sharper definition can remove it, because choosing the referent is itself a political judgment. The paper is explicit about what the method cannot yet do, including the human validation it still lacks, and the rubric, the prompt and the data are released in full. An early grounded pilot shows the size of that gap: re-scoring the same cells from cited documents moves a single model's placements more than switching models does. The worked example is the Middle East and North Africa, and in my estimation the method carries to any region the standard tools leave out.
\end{abstract}
\vspace{1.2em}
}]
\begingroup
\renewcommand{\thefootnote}{}
\footnotetext{Independent researcher. Correspondence via \texttt{tarekgara.com/tayyar}. The author is responsible for all judgments, scores and errors. The rubric and the versioned data releases are public at the address above. This version corrects an import error that had attached 154 of GPT's 961 ratings to the wrong parties (Appendix~\ref{app:rounds}). With those ratings reassigned, the panel covers 80 parties, where version 1 reported 89; the ordinal $\alpha$ is 0.87 (version 1: 0.86); and the per-model, per-axis and provenance tables move by a few hundredths. The provenance contrast on sectarian power-sharing rises from $+0.07$ to $+0.18$, on eleven cells, and the rubric replication now covers eight raters ($2.43 \rightarrow 2.12$; version 1: $2.07$). The description of the re-coding probe is also corrected. Every figure is recomputed from the corrected v1.0 data release.}
\endgroup

\section{Introduction}

Ask where a Lebanese party sits relative to an Egyptian one, or whether Tunisia's Ennahda in 2024
stands where Egypt's Muslim Brotherhood stood in 2012, and the tools built for this kind of comparison
mostly stop at the edge of Europe. The spatial tradition that treats politics as movement along
defined dimensions is old and well established~\cite{downs1957,hotelling1929,poole1997}, and the
manifesto-coding and expert-survey programs that put it to work are among the most cited tools in
the field~\cite{budge2001,bakker2015}. They were built for a different region. Many of the most
important actors in the Middle East and North Africa never publish a Western-style manifesto, never
contest a competitive election, or commit one position to a founding charter and the opposite one in
office. The result is a gap in coverage that goes well beyond a few missing rows.

This gap matters beyond area studies. Quantitative comparative politics increasingly uses position
estimates as variables in models of coalition formation, party competition, conflict onset and
democratic backsliding, so the narrow reach of these tools limits what the field can say. A field
that can place every party in the Bundestag to a tenth of a point, while it cannot place the actors of
a twenty-state region at all, is in effect deciding which questions get asked.

The obvious move today is to point a large language model at the problem and read the answer off. A
fast-growing literature reports that frontier models match or outperform crowd workers, and sometimes
trained coders, on text-annotation tasks~\cite{gilardi2023,tornberg2023}, which makes it tempting to
treat the model as a measurement device that has simply read more than any human could. I consider
this the wrong stance. A model's fluent and confident placement of a party is one reading, produced by
one system, and it carries biases from a training corpus the analyst did not choose and cannot
inspect. If the model is treated as an oracle, those biases end up laundered into apparent fact. The
alternative is to treat it as a \emph{rater}, that is, a judge whose output is pooled with others and
whose disagreements are kept and reported. A rater of this kind is something an existing
methodology already knows how to handle. For forty years, expert surveys have built valid measures
out of fallible human judges by pooling them and studying their agreement~\cite{bakker2015,benoit2006}.
This paper puts the language model in the same role, with the same tools and the same skepticism, and
asks the questions that methodology is designed to answer. Do the raters agree? If they do, is it
because they are right or because they share a mistake?

The instrument I build to make this concrete, Tayyar, is a positional dataset of the Middle East and
North Africa. It places parties and public figures, current and historical, on a set of defined and
regionally anchored axes, using a panel of raters that is, for now, mostly language models. I use it
throughout as a worked example, though the contribution of the paper lies elsewhere, in a method and a
stance: language models used as raters in data-sparse measurement, with their reliability taken
seriously and kept apart from validity. The main design choices follow from that stance, namely
applicability rules, a \emph{declared}-versus-\emph{behavioral} lens, a synthesis that keeps
disagreement, and an inclusion gate on aggregation.

Several results follow from this design. First, I set out the model-as-rater stance and the panel
behind it, including the measurement reasons to keep a blank distinct from a zero and a stated
position distinct from a revealed one. Second, a \emph{built-in experiment} isolates the effect of the
rubric by holding a definition-free round fixed. Written definitions both move scores and improve
agreement between raters, and this measurable rubric effect rules out the simplest form of the
steering worry, according to which the rubric merely imposes an arbitrary direction on a compliant
model. Third, I report Krippendorff's $\alpha$ under both interval and ordinal metrics, so the result
does not depend on an arguable equal-spacing assumption, and I press the point that high agreement
among models trained on overlapping data falls short of validity. The \emph{pattern} of disagreement
turns out to be informative as well. The panel's widest split, on an actor's stance toward the state's
foundational order, traces to a referent ambiguity that no rubric can resolve, and coding errors
account for little of it. Throughout, I try to be explicit about what the instrument cannot yet do, so
that a reader can weigh it fairly.

If the stance holds, the field gains a measurement where it currently has none: the actors of a
data-sparse region placed on comparable, defined axes, with their disagreement reported, by raters
that are available at scale today. The design is organized around its own central risk, that capable
models can be capably wrong in the same direction, and it tries to measure that risk and keep it in
view, leaving room for the human validation that would close it.

It is worth noting at the outset that the reliability figures below cannot stand in for an external
check. The usual external benchmarks, such as a manifesto corpus, an expert survey like the Chapel
Hill placements or party-level democracy scores, barely reach these actors, which is the very gap the
method exists to address. The validity test therefore has to come from purpose-built human anchoring.

\section{Related work}\label{sec:related}

Tayyar sits where several literatures meet, and it tries to borrow the discipline of each while
avoiding the blind spot that tends to come with it. I take the traditions one at a time, since most of
the method's design choices are easiest to understand as responses to a specific limitation in each.

\paragraph{Spatial models of politics.} The idea that political actors can be located as points in a low-dimensional issue space runs from Hotelling and Downs~\cite{hotelling1929,downs1957} through the roll-call scaling of Poole and Rosenthal~\cite{poole1997}, and it remains the common language of quantitative comparison. Tayyar keeps the geometry and drops an assumption that suits two-party legislatures far better than this region, namely that a single recovered dimension carries most of the signal. In the Middle East and North Africa, a party's stance on the role of religion in the state, its position on the regional axis of resistance versus normalization, and its posture toward the Palestinian question are empirically distinct, and they do not fold back into a left--right line. Collapsing them too early discards the structure the measurement was meant to recover. The instrument hence keeps a full position vector and treats the familiar two-axis ``compass'' as a default view of it.

\paragraph{Manifesto coding and text scaling.} The Comparative Manifesto Project and its MARPOR successor~\cite{budge2001,volkens2021,klingemann2006} established the practice of estimating positions from a party's own published texts, hand-coded sentence by sentence against a fixed scheme. Its automated descendants, ``words as data''~\cite{laver2003}, the Wordfish scaling model~\cite{slapin2008} and the later refinements of scaling from coded text~\cite{lowe2011}, dropped the human coder at the price of needing a large reference corpus in a language the method has seen at scale. Two points matter here. First, the coding is noisier than its wide use suggests. Mikhaylov, Laver and Benoit show non-trivial misclassification even among trained human coders of manifestos~\cite{mikhaylov2012}, a sobering baseline for any automated substitute and a reminder that the human coder is one more rater, with no claim to being a gold standard. Second, and decisively for this setting, bag-of-words scaling is tied to its corpus and its language. Arabic is comparatively under-served by the pretrained representations these methods rely on~\cite{antoun2020}, and Hebrew more so. A Wordfish estimate comparing an Arabic charter to a Hebrew platform compares two thinly modeled languages through a method that assumes a dense one. Grimmer and Stewart are candid that automated text methods amplify careful reading and cannot replace it~\cite{grimmer2013,grimmer2022}. Tayyar keeps document grounding as the spine of its \emph{declared} reading and replaces word-frequency scaling with rubric-anchored judgment, which a multilingual model applies more evenly across Arabic, Hebrew and English than a model of one language's word counts can.

\paragraph{Expert surveys and the aggregation of judgment.} The Chapel Hill Expert Survey~\cite{bakker2015,hooghe2002} and Benoit and Laver's expert placements~\cite{benoit2006} ask informed people to place parties directly and take the mean expert as the estimate. Their validity rests on inter-expert agreement and on the expectation that idiosyncratic error washes out across a panel. The Varieties of Democracy project goes furthest in this direction, aggregating many country experts through an explicit item-response measurement model that estimates both the latent positions and each rater's reliability and thresholds~\cite{vdem2024,pemstein2018}. This is the tradition Tayyar belongs to most closely, and the analogy is close to literal, since the panel is an expert survey in which most of the ``experts'' are language models. It inherits the tradition's main strength, a principled way to build a measure out of fallible judges, together with the vulnerability V-Dem's own methodologists take care over. A panel that agrees for a shared wrong reason looks, from inside the agreement statistics, exactly like one that agrees because it is right. Human expert panels are assembled to be diverse, whereas a panel of language models can be far more correlated than its size suggests, a point I return to at length.

\paragraph{Cross-cultural incomparability.} Placing actors from very different systems on one scale raises the problem that the scale may not mean the same thing in each. King and colleagues formalized this as differential item functioning and proposed anchoring vignettes to correct it~\cite{king2004}. The Aldrich--McKelvey family of methods and its Bayesian extensions recover latent positions while estimating and removing each rater's distortion of the common scale~\cite{aldrich1977,hare2015}. For Tayyar the problem is live in two ways. Across actors, an axis can simply fail to be commensurable. ``West alignment'' names an existential strategic baseline for an Israeli party and a contested ideological choice for an Arab one, so the same number encodes different facts on the two sides of that divide. Across raters, different models may anchor the $-10$ to $+10$ scale in different places. The approach here manages incomparability by anchoring the rubric in worked regional examples that fix what the poles mean, by scoping axes to where they apply, and by flagging the axes whose comparability across the divide is weakest. These defenses are partial, and Section~\ref{sec:limitations} returns to what they leave open.

\paragraph{Language models as annotators.} The neighboring literature finds that frontier models are strong text annotators. Gilardi, Alizadeh and Kubli report ChatGPT outperforming crowd workers on several political-text tasks~\cite{gilardi2023}, T\"ornberg reports much the same for annotating the ideology of political messages~\cite{tornberg2023}, and Ziems and colleagues survey the broader promise for computational social science while cataloging where it can go wrong~\cite{ziems2024}. A parallel strand insists that LLM annotations be \emph{validated} against human labels and shows that unvalidated automated coding can be confidently wrong~\cite{pangakis2023}. Tayyar takes both messages seriously. It uses models as annotators because, in a region where trained coders and reference corpora are scarce, they are the most capable raters available at scale, and it builds validation in as planned human anchoring.

\paragraph{Language models as position estimators.} Closest to this paper is a young literature that uses language models to place actors directly in a policy or ideological space. Wu, Nagler, Tucker and Messing scale US senators on a liberal--conservative dimension by asking a model for pairwise comparisons, and recover a scale that correlates strongly with NOMINATE~\cite{wu2023}. Le Mens and Gallego take the approach furthest. They ask an instruction-tuned model where a text stands on a focal dimension and \emph{average} its responses, recovering manifesto and roll-call benchmarks at correlations above $0.9$ across several languages, and they warn that the method has to be re-validated in each new setting~\cite{lemens2025}. Tayyar shares the premise that a model can read a position and departs from this work in four places, which the rest of the paper develops. First, it treats the models as a \emph{panel} of distinct raters and keeps their disagreement, since the disagreement is itself diagnostic (Section~\ref{sec:reliability}). Second, it works in a region with no large validated reference corpus, so the benchmark correlations that make averaging look safe in well-served party systems are unavailable, and reliability has to be argued from the panel itself, with the threat of a shared prior addressed directly. Third, it gates applicability and declines to score an axis that does not apply. Fourth, it takes Le Mens and Gallego's warning, that a high correlation or a high $\alpha$ can hide a shared error, as its organizing idea, and places the distinction between reliability and validity at the center of the method.

\paragraph{The political leanings and non-independence of models.} A panel of models cannot be read like a panel of independent experts, for the simple reason that models are not independent. A growing body of work documents that language models carry measurable, and often shared, political tendencies. Santurkar and colleagues show whose opinions models reflect and whose they leave out~\cite{santurkar2023}, Durmus and colleagues measure how some national viewpoints are over-represented on global-opinion tasks~\cite{durmus2024}, Atari and colleagues argue that the representations skew toward Western, educated and industrialized populations~\cite{atari2023}, and Feng and colleagues trace political bias from pretraining data through to downstream behavior~\cite{feng2023}. Several studies find a consistent left-of-center tilt across widely used models~\cite{rozado2024,motoki2024}, with the deeper diagnosis that the training corpora, and the people who built them, shape the prior~\cite{bender2021}. A related literature on ``silicon sampling'', which uses models as stand-ins for survey respondents, finds the simulations convincing and unreliable at once, faithful to some subpopulations and badly off for others~\cite{argyle2023,bisbee2024}. Is a model therefore a useless rater? It seems not, though its agreement is suspect in a specific and measurable way. Models drawn from overlapping corpora may share a prior, and about a region like the Middle East that prior is plausibly an Anglophone-media one. The reliability analysis of Section~\ref{sec:reliability} confronts this threat directly, and only the planned human anchoring can settle it.

\paragraph{Reliability theory.} The standard coefficients are Cohen's $\kappa$ for chance-corrected categorical agreement~\cite{cohen1960} and, for ordered or interval scales on which not every rater scores every item, Krippendorff's $\alpha$~\cite{krippendorff2019,hayes2007}, which handles the missing data a partially filled panel always produces. I lead with $\alpha$, and following the advice to report a coefficient under the measurement assumption one is prepared to defend, I report it under both an interval and an ordinal difference function.

\paragraph{Regional scholarship.} The frame and its exemplars draw on the area-studies literature, including Lynch on the post-2011 uprisings~\cite{lynch2016}, Hinnebusch on Ba'athist Syria~\cite{hinnebusch2001} and Wickham on the evolution of the Muslim Brotherhood~\cite{wickham2013}. Anchoring the axes in this literature lets the rubric carry the meaning of the scale, which would otherwise default to each rater's own folk theory of the region.

\section{The instrument: frame, applicability, and lenses}\label{sec:frame}

\subsection{Axes}
The dataset runs on sixteen active axes (Table~\ref{tab:axes}). Each has two named poles, a written definition that governs how it is scored, and a scope that decides where it applies at all. Eight are general dimensions of political contestation. The others are regional or context axes, meant to catch cleavages a Western frame misses or misreads, such as pan-Arabism, the posture toward Iran and the stance on confessional power-sharing. The default two-axis compass uses the economic and social axes, to stay close to the political-compass tradition, while the clustering and comparison analyses use the full sixteen-dimensional vector. This is one defensible way to divide the region's politics among several, and Section~\ref{sec:limitations} returns to what the choice leaves out.

\begin{table*}[t]
\centering\footnotesize
\setlength{\tabcolsep}{5pt}
\begin{tabular}{@{}lll@{\hspace{1.6em}}lll@{}}
\toprule
\textbf{Axis} & \textbf{$-$ pole} & \textbf{$+$ pole} & \textbf{Axis} & \textbf{$-$ pole} & \textbf{$+$ pole} \\
\midrule
Economic & Statist & Market & Regime stance & Anti-regime & Pro-regime \\
Social & Authority & Libertarian & Pan-Arabism & Particularist & Pan-Arab \\
State \& religion & Religious state & Secular state & Federalism & Centralist & Federalist \\
Liberal democracy & Weak / anti & Strong commitment & Modernization & Traditionalist & Modernizing \\
West alignment & Anti-Western & Pro-Western & Gender equality & Patriarchal traditionalism & Gender equality \\
Regional stance & Resistance / maximalist & Stability / normalization & Iran posture & Anti-Iran / adversarial & Pro-Iran / aligned \\
Palestinian question & Opposed & Pro-Palestinian rights & Press freedom & State-controlled press & Free press \\
Civil liberties & Restrict & Expand & Sectarian power-sharing & Consociational / quota & Post-sectarian / civic \\
\bottomrule
\end{tabular}
\caption{\textbf{The axis frame.} Each of the sixteen active axes carries a written definition and
worked regional anchors that fix its poles; the definitions govern scoring, and the labels only name the poles.}
\label{tab:axes}
\end{table*}

\subsection{Applicability, and why a blank is not a zero}
Most axes are universal, though a handful are scoped, and the handling of scope turns out to be a measurement decision with consequences. Pan-Arabism is simply not a meaningful question for a Jewish-Israeli party. The stance on confessional power-sharing applies only where such a system exists, as in Lebanon's Taif arrangement or Iraq's \emph{muhasasa}, and context axes apply only to their countries. When an axis does not apply to an actor, Tayyar leaves the cell \emph{blank}, and both scoring and rendering skip it.

In a spatial model, a score of zero places the actor at the midpoint of a contested dimension, whereas an inapplicable axis means the actor is not on the dimension at all. Coding the second as the first injects a phantom centrist position that drags every aggregate over the axis toward the middle. A region's mean ``sectarianism'' score, with the inapplicable cells set to zero, ends up measuring mostly how many actors the question does not concern. Being outside a question and being at its center are different facts, and a measurement that conflates them misleads any downstream model that averages, clusters or correlates over the axis. In a region whose cleavages reach so unevenly, I see no workable alternative to keeping the blank distinct from the zero.

\subsection{The lens system: declared, behavioral, composite}
For many political actors, a single number papers over the gap between what they profess and what they do. For this reason, Tayyar models a position as up to three readings. The \emph{declared} lens is the stated position, read from platform, charter and dated speech, and it is the reproducible spine of the measurement, since two raters reading the same document can be asked to converge. The \emph{behavioral} lens is what the actor does with power, meaning its legislative record and its conduct in office, and it serves to test the declared position. The \emph{composite} is the published synthesis.

Each cell is assigned to the construct that fits it. The declared lens is the default. The behavioral lens is scored \emph{selectively}, for actors who hold or have held power, on the conduct axes of liberal democracy, civil liberties and press freedom, where a manifesto promise is cheap and the governing record is the better evidence. A record in office speaks to these three axes more directly than to the others. How a party treats courts, dissent and the press shows in its conduct in a way its pan-Arab sympathies or its abstract stance toward the regime's legitimacy do not, and an actor that has never governed leaves no such record, so extending the behavioral lens to the other axes would mostly manufacture missing data. The approach borrows from the behavioral turn that distinguishes what regimes say from what they do~\cite{vdem2024}. An actor whose character changed across eras is modeled as separate era-stamped entities, so that one cell never has to hold the contradiction; the founding 1947 Ba'ath of Aflaq's pan-Arab charter is a different entity from the Assad regime that governed in its name. The split also removes a tension that would otherwise corrupt the reliability statistics. A good share of the cells on which the panel disagrees most turned out to be declared and behavioral readings that an undifferentiated score had been averaging together, and separating the constructs turns that spurious disagreement into two well-posed measurements.

\section{Data and provenance}\label{sec:data}

\begin{table}[t]
\centering\small
\begin{tabular}{@{}lr@{}}
\toprule
\textbf{Layer} & \textbf{Count} \\
\midrule
Countries in the frame & 20 \\
Political parties (current and historical) & 98 \\
Public figures & 274 \\
Active axes & 16 \\
Source documents (platforms, charters, speeches) & 425 \\
Bills and legal instruments & 61 \\
Timeline events & 353 \\
Verified attributed quotes & 101 \\
Panel ratings recorded & 12{,}867 \\
\bottomrule
\end{tabular}
\caption{\textbf{Coverage in the v1.0 release.} Parties are scored in fifteen of the twenty framed
countries; five Gulf states have no parties on record.}
\label{tab:coverage}
\end{table}

The unit of analysis is the party or the public figure. Current and dissolved actors are both included and era-stamped, so historical comparison is supported directly. Scored positions are grounded, where grounding exists, in primary documents such as party platforms and charters, dated speeches, official statements and founding constitutions, while secondary sources serve only to corroborate biographical facts. For non-English originals, each source records the language it was written in, a detail that matters because a model's reading of a translated text and of the original need not agree. Verification is explicit and runs through three states, \texttt{unverified}, \texttt{tarek-verified} and \texttt{externally-verified}, and every record carries its state.

The provenance bears on how the reliability evidence should be read, so I state it plainly. The dataset is largely the work of one principal investigator with model assistance. The verification ladder, the document-grounding layer and the planned human anchoring exist to keep that provenance legible and improvable. A reader should, in my view, treat the current dataset as a well-instrumented pilot, whose internal agreement is real and whose external validation has not yet been done.

\section{The rater panel and its synthesis}\label{sec:panel}

\subsection{The panel}
A position in Tayyar comes out of a rater panel. The choice everything else depends on is to treat each language model the way an expert survey treats a single expert~\cite{bakker2015,benoit2006,pemstein2018}, as a judge whose worth shows once the judgments are pooled and studied, and who can simply be wrong. The analogy breaks down in one place that matters, and I return to it in Section~\ref{sec:reliability}. A human expert panel is assembled for a spread of training and background. These raters share web-scale corpora, and two of them, Gemini and Gemma, come from the same builder. Whatever epistemic diversity the panel has is thus something to test, and nothing in its composition guarantees it.

Each cell can carry ratings from up to four kinds of rater. An \emph{author-baseline} hand-coding seeds the panel as an explicit prior; it is the one rating in the system that does not come from a language model, and the panel is tracked against it. The \emph{language models} number nine: Claude, Gemini, Grok, GPT and Gemma, built in the United States, and Kimi, DeepSeek, Qwen and MiniMax, built in China, drawn from eight laboratories (Appendix~\ref{app:roster} lists the version each model ran). Each scores every cell on its own, in a fresh session, from a blind template that shows the axis definition and poles and none of the other raters' answers. Models join as they become available. A model that joins partway through scores from that point on, and both the synthesis and the reliability statistics tolerate a rater with partial coverage. A tenth model was trialed and dropped because it could not complete the scoring instrument reliably. \emph{Human anchors}, scored by people, are a planned round whose purpose is the validity check the model panel cannot run on itself (Section~\ref{sec:reliability}); they are not in the data yet. Two smaller layers complete the picture. In a \emph{grounded pilot}, one model scored directly from the cited documents, and in an early \emph{re-coding probe}, two language-model re-coders scored a sample of 63 cells. Appendix~\ref{app:rounds} describes both.

\subsection{Synthesis, and the preservation of disagreement}
Each rater's ratings for a cell collapse to that rater's median, and the published score is the median \emph{across distinct raters}, so a model that scored a cell in three rounds counts once. I use the median because, in a panel of nine, one model that badly misreads a cell should not be able to drag the published number, and the median, which resists outliers, is the choice an expert-survey methodology would make in any case. The published median draws on the with-definitions model round (r2), the author's prior, the grounded pilot, which is folded into Claude's value so that Claude still counts once, and, on their 63 cells, the two re-coders. The definition-free rounds never enter it, since they are the baselines held back for the experiment of Section~\ref{sec:rubric}.

The synthesis keeps disagreement visible. Dispersion is summarized by the median absolute deviation (MAD) of the raters' scores about the published median, on the $-10$ to $+10$ scale, and each cell receives a status from fixed cutoffs. A cell is \emph{settled} when it has at least two raters whose MAD about the median is within tolerance (MAD~$\le 1.5$). It is \emph{contested} when they diverge sharply (MAD~$\ge 3$), in which case the cell is rendered and released as a range. It is \emph{provisional} otherwise, which covers a single rater so far or a spread between the cutoffs. The thresholds are pragmatic, and the underlying MAD travels with each cell so that a reader can redraw them. A companion agreement figure, $1 - \mathrm{MAD}/10$, accompanies every synthesized cell as a convenience for the live tool; it rescales dispersion onto $[0,1]$ for display and has no statistical standing of its own. The v1.0 release holds 12{,}867 ratings. Of its 1{,}108 active composite cells, 803 (72\%) are settled, 53 (5\%) are contested and 252 (23\%) remain provisional, and 998 (90\%) carry at least two independent model raters. The panel fills country by country, so these proportions shift with each scoring session and should be read as interim.

\subsection{An inclusion gate on aggregation}
A cell's status governs that cell and nothing more. Field-level quantities, such as which axis divides the region most, how two party families compare, or how the axes correlate, are a second layer of inference, and a sparse axis can quietly distort all of them. With only a handful of actors scored on an axis, a single outlier swings the mean, a high variance reads as ``contested'' when the axis is merely thin, and a spurious correlation can end up at the top of a table. To guard against this, Tayyar admits an axis into field-level aggregates only once its coverage clears a floor, currently thirty scored entities. The figure is a convention I chose, though it falls in an empty band of the coverage distribution. No axis has between roughly twenty and sixty-five scored cells (its per-axis $n$ in Table~\ref{tab:axisrel}), so the split between core and periphery comes out the same for any floor in that range. Coverage is the structural gate, because coverage is what gives an average its meaning. Cross-rater dispersion is reported alongside it as a separate signal, so a well-covered and genuinely contested axis is flagged and stays eligible. Axes below the floor are reported in full on their own and never folded into a headline. In the current data, this separates a core of well-covered dimensions from a periphery of scoped axes (sectarian power-sharing, Iran posture, press freedom and gender) whose small cell counts make them informative about individual actors and not yet about the field. The gate is unglamorous, and I doubt the dataset would work without it, since most of the ways a positional dataset misleads at the macro level come down to reading thin coverage as a finding.

\section{The rubric effect: a built-in experiment}\label{sec:rubric}

The standard objection to a language model as a rater is that the rubric does the work. Hand a compliant model a definition with a built-in direction and it will echo that direction back, so any apparent agreement measures how suggestive the prompt was. The objection is fair, and the round protocol is designed to answer it with a number.

Ratings are labeled by round. The first round (r1) ran on two models, Gemini and Grok, \emph{without} the axis definitions. Each model received the poles and placed the actor from its own understanding, and the round was then locked. Every later round runs \emph{with} the sharpened definitions and the construct rule of Section~\ref{sec:frame}. Holding r1 fixed makes the worry measurable. The movement from r1 to r2 on the cells scored in both is the effect of introducing the rubric, and the shape of that movement shows whether the rubric steered the raters or corrected them.

Two findings come out of this. The first concerns movement within each rater. Across the 376 rater-cell pairs scored in both rounds, the mean absolute shift is 1.79 points on the 21-point scale, and r1 and r2 scores correlate at $r \approx 0.90$ (Figure~\ref{fig:rubric}a). In other words, the rubric leaves the central tendency of most placements in place, as one would expect if the models started from reasonable priors, but it matters in the tail, where roughly one cell in five (19\%) moves by three points or more. On inspection, the large moves read as corrections. In a fairly typical case, a model placed Likud at $+4$ on the liberal-democracy axis without definitions, taking ``holds elections'' to mean ``committed to liberal democracy,'' and at $-3$ once the rubric specified that the axis concerns the \emph{conduct} of power as well as the holding of elections. The move brought an outlier back in line with the rest of the panel and with the behavioral record.

The second finding is that the rubric improves agreement \emph{between} raters. On the 187 cells both models scored in both rounds, the definitions narrowed the mean absolute gap between Gemini and Grok from 2.81 to 2.50 points and raised their correlation from 0.81 to 0.89 (Figure~\ref{fig:rubric}b). The steering objection struggles with this. A rubric that imposed an arbitrary direction would shift the scores without necessarily bringing two independent raters closer about where an actor sits, and the definitions did both, which rules out a purely arbitrary or orthogonal steer. On its own, the result still cannot separate convergence toward correctness from convergence toward an anchor the rubric supplies. Only an external record, such as the behavioral evidence in the Likud correction, and eventually the planned human-anchor round can make that separation. Evidently, the experiment shows something narrower than one might want, and it holds: the first-order effect of the rubric is to turn vaguer individual readings into a more reproducible shared one.

Two raters are a thin base for a general claim, and the feature that makes the test clean, the \emph{same} two models scoring the same cells with and without definitions, is also its limit. A larger replication, paired less perfectly, helps with generality. The definition-free round was run across the panel on a fixed, randomly drawn anchor of 204 cells (the persistent inter-rater sample) and compared with the rubric round on the same cells. Mean pairwise disagreement between raters falls from 2.43 to 2.12 points, in the same direction and of comparable size, now across eight raters from seven laboratories. GPT is left out of this comparison, because the import of its definition-free ratings returned entity keys that did not match its rows (Appendix~\ref{app:rounds}). The convergence does not depend on which two models happened to run the first round.

The design carries beyond this dataset. Any rubric-based use of a model rater can run the same fixed, definition-free baseline and report what its rubric does.

\begin{figure*}[t]
\centering
\begin{tikzpicture}[x=0.82cm,y=0.021cm]
  \node[anchor=west,font=\small\bfseries] at (-0.6,168) {(a) Within-rater shift, r1 to r2};
  \draw[gray!55] (-0.6,0) -- (6.6,0);
  \foreach \x/\n in {0/98,1/135,2/71}{
    \fill[teal!35] (\x-0.36,0) rectangle (\x+0.36,\n);
    \node[font=\scriptsize,above] at (\x,\n) {\n};}
  \foreach \x/\n in {3/22,4/16,5/8,6/26}{
    \fill[teal!75!black] (\x-0.36,0) rectangle (\x+0.36,\n);
    \node[font=\scriptsize,above] at (\x,\n) {\n};}
  \foreach \x/\l in {0/0,1/1,2/2,3/3,4/4,5/5,6/6+}{\node[font=\scriptsize,below] at (\x,0) {\l};}
  \node[font=\scriptsize,below] at (3,-14) {absolute shift (points, rounded down)};
  \node[font=\scriptsize,align=center] at (4.5,72) {19\% move\\three points or more};
\end{tikzpicture}\hspace{2.2em}%
\begin{tikzpicture}[x=3.1cm,y=3.4cm]
  \node[anchor=west,font=\small\bfseries] at (-0.42,1.0) {(b) Mean gap between raters};
  \draw[gray!55] (0,0) -- (0,0.88); \draw[gray!55] (1,0) -- (1,0.88);
  \node[font=\scriptsize,below] at (0,0) {no definitions};
  \node[font=\scriptsize,below] at (1,0) {with definitions};
  \draw[thick,teal!75!black] (0,0.81) -- (1,0.50);
  \fill[teal!75!black] (0,0.81) circle (1.6pt); \fill[teal!75!black] (1,0.50) circle (1.6pt);
  \node[font=\scriptsize,left] at (0,0.81) {2.81};
  \node[font=\scriptsize,right,align=left] at (1,0.50) {2.50\quad Gemini--Grok,\\\phantom{2.50}\quad 187 cells};
  \draw[thick,teal!40] (0,0.43) -- (1,0.12);
  \fill[teal!40] (0,0.43) circle (1.6pt); \fill[teal!40] (1,0.12) circle (1.6pt);
  \node[font=\scriptsize,left] at (0,0.43) {2.43};
  \node[font=\scriptsize,right,align=left] at (1,0.12) {2.12\quad eight raters,\\\phantom{2.12}\quad 204-cell anchor};
\end{tikzpicture}
\caption{\textbf{The rubric effect.} (a) Absolute change in each rater's score when the axis
definitions were added, for the 376 cells that Gemini and Grok scored in both rounds. Most placements
move by less than two points; the 19\% that move by three or more (dark bars) read as corrections on
inspection, as in the Likud case in the text. (b) Mean absolute gap between raters on the same cells
without and with the definitions, for the original pair and for the larger replication on the
204-cell anchor. Both gaps narrow by about a third of a point.}
\label{fig:rubric}
\end{figure*}

\section{Reliability}\label{sec:reliability}

The reliability question for a panel instrument is whether the same frame, applied by different raters, lands in the same place. The figures here cover the 80 parties scored so far, across fifteen countries; positions for individual public figures come in a later round. They are interim and will move as the panel fills, so they describe a measurement still in progress.

\subsection{Agreement across the model panel}
On the 998 cells that carry at least two independent, blind model scores, the panel agrees closely. Krippendorff's $\alpha$ across the model raters is \textbf{0.86} on an interval difference function. As in the synthesis (Section~\ref{sec:panel}), each rater contributes one value per cell, its within-rater median, so the coefficient is computed over distinct raters and is not inflated by repeated rounds from the same model. The steps of the $-10$ to $+10$ scale may not be equally spaced, since whether the distance from $-2$ to $-1$ equals the distance from $0$ to $+1$ is an assumption. I also report $\alpha$ on an ordinal difference function, which assumes only that the values are ordered and weights a disagreement by how much of the observed distribution lies between the two ranks. The ordinal $\alpha$ is \textbf{0.87}, essentially the same as the interval figure. Because the two agree, the reliability claim does not depend on treating the scale as evenly spaced, the assumption much interval reliability reporting leaves unstated and the one a methods reviewer is most entitled to question. The interval coefficient held at 0.86 as the panel grew from five raters to nine, which offers some comfort against one narrow worry, that the early figure was an artifact of a few near-identical raters. It cannot speak to a prior shared across all nine, which a wider panel would simply carry along.

Each model's average agreement with the others is high (Table~\ref{tab:permodel}). Every model's mean correlation with the rest of the panel lies between $0.85$ and $0.90$, with mean absolute gaps around two points. The panel as a whole also tracks the author's baseline, the one comparison available today that is not itself a language model. That tracking is worth only so much. The baseline is my own hand-coding, written by the same person who wrote the axes, the anchors and the rubric the models receive, so agreement with it is what one would see if the rubric had simply passed my prior on to the models. It is a coherence check on the instrument, and external validation can only come from raters independent of the frame.\footnote{A separate internal probe had two language-model re-coders score a stratified sample of 63 cells, drawn as three blind rounds of 21 (Appendix~\ref{app:rounds}). Their agreement with each other is $\alpha \approx 0.95$. Both re-coders apply the same framework, which reproduces a score almost by construction, so the figure is a consistency floor and I do not treat it as a reliability result; the relative pattern across axes is the useful part.}

\begin{table}[t]
\centering\footnotesize
\setlength{\tabcolsep}{4pt}
\begin{tabular}{@{}lllrrr@{}}
\toprule
\textbf{Model} & \textbf{Built} & \textbf{Laboratory} & \textbf{Cells} & \textbf{$\bar r$} & \textbf{mean $|\Delta|$} \\
\midrule
Claude   & US    & Anthropic & 997     & 0.90 & 1.88 \\
GPT      & US    & OpenAI    & 961     & 0.89 & 1.89 \\
Gemma    & US    & Google    & 1{,}001 & 0.87 & 2.21 \\
Gemini   & US    & Google    & 997     & 0.87 & 2.29 \\
MiniMax  & China & MiniMax   & 974     & 0.86 & 2.17 \\
Kimi     & China & Moonshot  & 997     & 0.86 & 2.19 \\
DeepSeek & China & DeepSeek  & 934     & 0.86 & 2.25 \\
Qwen     & China & Alibaba   & 997     & 0.85 & 2.32 \\
Grok     & US    & xAI       & 997     & 0.85 & 2.33 \\
\bottomrule
\end{tabular}
\caption{\textbf{Per-model agreement.} Each model's agreement with the rest of the panel on the r2
cells, ordered by mean pairwise correlation. $\bar r$ is its mean Pearson correlation with each other
model over shared cells, mean $|\Delta|$ the matching mean absolute gap on the $-10$ to $+10$ scale,
and ``Cells'' its r2 coverage. Every $\bar r$ falls in $0.85$--$0.90$, and build country does not order
the column: the highest and the lowest are both US-built, with the China-built models between them.}
\label{tab:permodel}
\end{table}

\subsection{Where the panel converges, and where it frays}
Agreement is far from uniform across the frame, and the per-axis pattern (Table~\ref{tab:axisrel}) is the actionable part, since it points to the rubric anchors that most need tightening. The panel converges most where a position is loudly and repeatedly declared, as on gender, the Palestinian question, and the social and state-and-religion dimensions, where two readers of the same texts land in the same place. Iran posture and press freedom look comparably tight in Table~\ref{tab:axisrel}, though they rest on thin coverage ($n=17$ and $n=19$), and I would not lean on their agreement yet. The reasons the panel frays elsewhere deserve naming, mostly because they differ from the obvious guess.

Take the single widest disagreement in the data. On regime stance, Hayat Tahrir al-Sham draws scores across the \emph{entire} scale (Figure~\ref{fig:referent}). Five models place it near the revolutionary floor (three at $-10$, two at $-9$) and four near the incumbent ceiling (between $+7$ and $+8$). This is no coding error, and a knowledge gap explains it only in part. In the eighteen months before this writing, HTS went from the most heavily sanctioned jihadist franchise in the Levant to the government in Damascus, and one might guess that the split separates the models that learned of the transition from those that did not. The rationales show otherwise. Raters at \emph{both} poles state the same facts, that HTS toppled the Ba'athist order and now governs Syria, and they divide on \emph{which} order ``the foundational order'' denotes. A rater scoring the actor's posture toward the order it overthrew reads it as anti-regime ($-10$), and a rater scoring its posture toward the order it is now building reads it as pro-regime ($+8$). I call this a \emph{referent problem}. The axis assumes a single stable order to take a stance toward, and an actor that has just exchanged one order for another breaks that assumption. Ennahda splits the same way across fifteen points, opposing Kais Saied's incumbent order while defending the 2011 constitutional order it helped write, and so do the Syrian Ba'ath, the Sudanese Communist Party and Yemen's Islah, each an actor whose country offers more than one plausible ``regime'' to be for or against. A sharper version is referent \emph{absence}. In a state mid-collapse, with no settled order to support or oppose, the axis is ill-posed. The applicability rule (Section~\ref{sec:frame}) asks whether an axis applies to an actor in principle and does not check whether a stable referent exists at a given moment, so it misses this case. Leaving such cells blank on the order-dependent axes, which the rule currently renders as a contested range, is a refinement it should absorb.

To test whether this is the mechanism, three language-model coders, working blind and independently, classified the source of disagreement on the twenty-two most divided regime-stance cells, agreeing on 76\% of their pairwise judgments. A clear majority traced to \emph{interpretation}. Roughly two-thirds (64\%) turned on which order is the referent, or on how to weigh an actor's posture toward it, against 18\% genuine factual conflict and 18\% definitional confusion. Written definitions are the rubric's instrument for the definitional kind, and on this axis they have little to work with. The panel divides over which regime an actor is scored against, and fixing that referent for a contested actor is a political judgment the instrument should not make. Accordingly, the synthesis marks these cells contested and keeps the insurgent and the incumbent readings apart, and the dataset flags the column to be read with that limit in mind.

This is the kind of finding I hoped the panel would catch. A single-number method would average the two clusters into a meaningless midpoint, while keeping the split exposes a genuine ambiguity in the question. For an actor that has just replaced one order with another, ``stance toward the regime'' has no single answer, and a confident zero would be the wrong one.

\begin{figure}[t]
\centering
\begin{tikzpicture}[x=0.335cm,y=1cm,score/.style={circle,fill=teal!70!black,inner sep=1.4pt}]
  \foreach \ry/\name in {1.4/{Hayat Tahrir al-Sham (Syria)}, 0/{Ennahda (Tunisia)}}{
    \draw[gray!45] (-10.4,\ry) -- (10.4,\ry);
    \foreach \t in {-10,-5,0,5,10}{\draw[gray!45] (\t,\ry+0.07)--(\t,\ry-0.07);}
    \node[anchor=south,font=\small\itshape] at (0,\ry+0.4) {\name};
  }
  % Hayat Tahrir al-Sham (y=1.4); coincident scores jittered vertically
  \node[score] at (-10,1.28){};\node[score] at (-10,1.4){};\node[score] at (-10,1.52){};
  \node[score] at (-9,1.34){};\node[score] at (-9,1.46){};
  \node[score] at (7,1.4){};
  \node[score] at (8,1.28){};\node[score] at (8,1.4){};\node[score] at (8,1.52){};
  % Ennahda (y=0)
  \node[score] at (-9,0){};\node[score] at (-8,0){};\node[score] at (-7,0){};
  \node[score] at (-5,0){};\node[score] at (-4,0){};
  \node[score] at (4,0){};\node[score] at (5,0){};
  \node[score] at (6,-0.06){};\node[score] at (6,0.06){};
  % pole labels
  \node[anchor=north west,font=\scriptsize] at (-10.4,-0.45){$-10$ anti-regime};
  \node[anchor=north,font=\scriptsize] at (0,-0.45){$0$};
  \node[anchor=north east,font=\scriptsize] at (10.4,-0.45){pro-regime $+10$};
\end{tikzpicture}
\caption{\textbf{The referent problem on regime stance.} Each dot is one of the nine models' scores
for the actor, with coincident scores jittered vertically. Both actors split into two clusters separated
by most of the scale, because the models agree on the facts and disagree on \emph{which} constitutional
order ``the regime'' denotes. For Hayat Tahrir al-Sham, the lower cluster scores its posture toward the
Ba'athist order it overthrew (anti-regime) and the upper cluster toward the post-Assad order it now
governs (pro-regime). For Ennahda, the split is between Kais Saied's incumbent order, which it opposes,
and the 2011 republican order, which it defends. Each cluster mixes US-built and China-built raters, so
the split follows interpretation and does not track build country. A blind triple-coding of the
twenty-two widest regime-stance cells attributes 64\% of such disagreement to interpretation of this
kind.}
\label{fig:referent}
\end{figure}

A second cluster sits on sectarian power-sharing, the loosest axis in the table, and it exposes a definitional trap. Hezbollah, the Amal Movement and the Syrian Ba'ath each draw spreads of thirteen to fourteen points, with raters scattered from pole to pole. The axis is meant to measure a \emph{stated position} on the confessional quota, that is, whether an actor would abolish \textit{muhasasa} or entrench it. A model that hears ``Hezbollah'' and scores a sociologically sectarian actor answers a different question from one that reads the party's own anti-confessional rhetoric and scores the platform. Both readings are defensible, and they answer different questions. The lens system (Section~\ref{sec:frame}) exists to resolve this kind of construct conflation.

The same pattern runs across the frame. The panel agrees about what actors \emph{say} and frays over what their position \emph{is} once it has to be inferred. This holds on the conduct axes (liberal democracy and civil liberties), where a charter and a record pull apart, and on the scoped regional axes (federalism, regime stance, pan-Arabism and above all sectarian power-sharing), where raters differ about what the axis asks before they reach an answer. The axes that define the region's conflicts are among the \emph{least} settled, which is roughly the diagnosis a reliability program should produce. It directs the next round of rubric work, and it tells the reader which columns of the dataset to lean on and which to treat with caution.

This creates a real tension. The panel is tightest on the universal, Western-legible axes, the social, economic and state-and-religion core, and loosest on the regional axes (pan-Arabism, sectarian power-sharing and regime stance) that justify its existence, so its reliability is lowest exactly where the instrument claims to add the most. Admittedly, this could be read as a refutation of the method. I read it as the diagnosis of a problem. The regional, evaluatively loaded constructs carry the interpretive disagreement analyzed above, and an instrument that reported them as settled would be hiding that problem. The implication for use holds either way, and the inclusion gate of Section~\ref{sec:panel} enforces it: the regional columns inform claims about individual actors and stay out of field-level aggregates until their coverage and agreement can support the weight.

\begin{table}[t]
\centering\small
\begin{tabular}{@{}lrr@{}}
\toprule
\textbf{Axis} & \textbf{n cells} & \textbf{mean $|\Delta|$} \\
\midrule
Palestinian question & 80 & 1.33 \\
Gender equality & 20 & 1.51 \\
Iran posture & 17 & 1.55 \\
Press freedom & 19 & 1.81 \\
State \& religion & 80 & 1.81 \\
Social & 80 & 1.82 \\
Modernization & 76 & 1.87 \\
West alignment & 79 & 1.99 \\
Economic & 80 & 2.00 \\
Civil liberties & 79 & 2.07 \\
Liberal democracy & 80 & 2.27 \\
Regional stance & 79 & 2.45 \\
Federalism & 76 & 2.70 \\
Regime stance & 76 & 2.90 \\
Pan-Arabism & 66 & 3.16 \\
Sectarian power-sharing & 11 & 4.03 \\
\bottomrule
\end{tabular}
\caption{\textbf{Per-axis agreement.} Mean pairwise absolute gap across the panel, tightest to
loosest, on the $-10$ to $+10$ scale. Scoped axes have small cell counts and are noisier, so each row
should be read with its $n$; the inclusion gate (Section~\ref{sec:panel}) keeps the thinnest out of
field-level aggregates.}
\label{tab:axisrel}
\end{table}

\subsection{Reliability is not validity}
Everything in this section concerns \emph{reliability}, the reproducibility of a reading, and nothing in it establishes \emph{validity}, its correctness. The distinction matters more than almost anything else in the method. Most of the panel's raters are language models trained on overlapping data, and nine models agreeing amount to nine correlated readings~\cite{ziems2024}. They can share a single prior, and for the politics of the Middle East, a prior shaped by Anglophone media coverage is exactly the kind of shared error I would most want to rule out. A high $\alpha$ is fully consistent with an instrument that is reliable and systematically wrong, and no statistic computed over the models alone can tell the two apart. The defenses lie outside the panel: the author's baseline that the panel tracks (a single non-model anchor, useful and insufficient alone), the document-grounding layer that ties scores to cited text, and the one that matters most and is not yet collected, human anchors scored by people.

One internal check makes the gap concrete. The archived grounded pilot, in which a single model scored from the cited documents, overlaps the parametric panel on 117 cells. There, its grounded scores correlate only $r = 0.65$ with its own parametric scores and differ from them by a mean of 3.7 points on the 21-point scale. The gap exceeds any model's mean disagreement with the rest of the panel (1.9 to 2.3 points in Table~\ref{tab:permodel}), and 44\% of the cells move by three points or more. Re-reading the documents moves the same model more than swapping one model for another does. This is one model on a small, early sample, and it settles neither reading, though it points the same way as the rest of this section. The panel's close agreement is agreement about a shared parametric reading, and grounding that reading in cited text is the validity work that remains, which adding more models cannot do.

One further check can run inside the panel. It mostly comes up empty, and I report it for the record. If the shared prior were specifically a US or Anglophone one, models built in the United States and models built in China should disagree wherever that prior drives the score more than the text does. I measure this per axis as the mean absolute gap between US-built and China-built raters on the cells both blocs scored (the \emph{between-bloc} gap), set against the mean gap among raters of the same build country (the \emph{within-bloc} gap, a five-rater quantity on the US side and a four-rater one on the China side). On no axis does the between-bloc gap exceed the within-bloc gap by more than a third of a point (Table~\ref{tab:provenance}). The largest difference is on regime stance ($2.74 \rightarrow 3.02$, $+0.29$), and the next, on federalism and Iran posture, never exceeds $+0.25$. Of the two axes a naive geopolitical-bias story would target first, pan-Arabism shows almost no difference, and sectarian power-sharing's $+0.18$ rests on eleven cells. The split is small and does not single out the China-built models.

I would not lean on this result heavily. Build country is a crude proxy for what actually matters, corpus independence: Google built two of the nine raters, ``US-built'' groups corpora that share little, and four raters are a thin base for a within-bloc dispersion. A null result here also cannot rule out a prior shared across \emph{all} nine models whatever their origin, which is the harder and probably the likelier worry. A fuller treatment of the provenance contrast is left to companion work. The test that would settle the question lies outside the models altogether, and the validity round that would supply it is the method's main unfinished business.

\begin{table}[t]
\centering\small
\begin{tabular}{@{}lrrr@{}}
\toprule
\textbf{Axis} & \textbf{within} & \textbf{between} & \textbf{difference} \\
\midrule
Regime stance           & 2.74 & 3.02 & $+0.29$ \\
Federalism              & 2.57 & 2.81 & $+0.25$ \\
Iran posture            & 1.42 & 1.65 & $+0.23$ \\
Civil liberties         & 1.96 & 2.16 & $+0.20$ \\
Sectarian power-sharing & 3.93 & 4.11 & $+0.18$ \\
Pan-Arabism             & 3.13 & 3.17 & $+0.04$ \\
\bottomrule
\end{tabular}
\caption{\textbf{The provenance contrast is small.} Mean absolute gap between raters within
build-country blocs and across them, on the $-10$ to $+10$ scale, for the five axes with the largest
between-minus-within difference and for pan-Arabism (the rest are smaller or negative). The largest,
on regime stance, is under a third of a point. Of the two axes a geopolitical-bias account would
target first, pan-Arabism shows almost none, and sectarian power-sharing rests on eleven cells. Build
country is a coarse proxy for corpus independence (see text). Differences are computed on unrounded
values and may differ from the rounded columns by~$0.01$.}
\label{tab:provenance}
\end{table}

\section{Limitations}\label{sec:limitations}

\begin{itemize}[leftmargin=1.2em]
\item \textbf{Correlated model error, and the missing validity check.} The panel consists mostly of language models trained on overlapping corpora, so their agreement is evidence of \emph{reliability} and says little about independence (Section~\ref{sec:reliability}). A contrast of models by national provenance, developed in companion work, tests whether a shared prior is doing the work, though detecting a bias is a different task from removing it. Human anchors, scored by people blind to the panel, would turn reliability into \emph{validity}. That round is planned and not yet collected, and until it is, the gap applies to every reliability figure in this paper. It will also close unevenly, for reasons built into the data. Human anchoring is feasible for present-day, reachable actors and hardest exactly where the data are thinnest, for defunct parties and for actors in active conflict zones where local expertise is scarce or unsafe to solicit. For those cases, the document-grounding layer, which ties a score to cited text and needs no living expert, is the validity path that has to carry the weight, so validity will arrive unevenly across the dataset.
\item \textbf{Stated versus revealed.} The \emph{declared}/\emph{behavioral} lens manages the oldest problem in positional measurement and does not make it disappear. Which object a number refers to is a choice the analyst makes and has to declare, and a reader who draws that line elsewhere should read some cells differently.
\item \textbf{Ungrounded model rounds, and why grounding is the more valuable work.} The current rounds score from the models' parametric knowledge plus the rubric, and only the pilot scores from cited documents. They are best read as ``what well-informed models believe, given clear definitions,'' with document-grounded scoring as a separate verification layer still under construction~\cite{laver2003,pangakis2023}. Reading parametric knowledge means the rounds may be extracting the shared prior itself, so the reliability statistic and the scoring construct can agree for the same reason, both reading the prior when they should be reading the actor. Grounding the score in cited text (Section~\ref{sec:reliability}) would close that gap, and it is the unfinished work I would put first.
\item \textbf{Recency, and the moving target.} Politics in the region moves fast, and a model's knowledge of the latest turns is bounded by its training cutoff. The prompt dates the question (``score the actor as it stands in 2026''), yet an actor mid-transition, such as Hayat Tahrir al-Sham between insurgency and government, or a party turned out of office between one model's cutoff and the next, can still split the panel across the full scale when the models disagree about a fact only some of them have learned. The design treats such splits as signal. The cells are flagged \emph{contested}, and the synthesis does not average an insurgent with an incumbent. The blind triple-coding of the widest regime-stance cells (Section~\ref{sec:reliability}) is one attempt to tell the two sources apart. It attributed 18\% of that disagreement to factual conflict, the category that would contain any cutoff-driven split, against 64\% to interpretation, so cutoff-driven splits exist there and are in the minority. Recency is a hard ceiling on any model panel, and the only real remedy is fresh scoring once events settle.
\item \textbf{The frame is one decomposition among several.} Every axis is a defensible choice and also an editorial call. The West-alignment axis in particular means different things for Israeli and Arab actors, so comparisons across that divide require care~\cite{king2004}. I do not yet report how much the headline placements move when a given axis is dropped or redefined, and a sensitivity analysis of that kind, which the regime-stance case (Section~\ref{sec:reliability}) argues for, is a necessary next step.
\item \textbf{An uneven, drifting panel.} Not every cell is scored by the same set of models. Per-model coverage varies, so the reliability figures pool across cells whose panels differ in composition. A commercial model is also a moving target, because a \emph{rater} label means whichever version answered in that session, and recording model identifiers only partly pins that drift down. The recording is itself incomplete, since the version field was left empty for several models' main-round sessions (Appendix~\ref{app:roster}).
\item \textbf{Imports.} Ratings returned from chat sessions were imported by key, and three of these imports came back with keys that did not match their rows. In the main round, 154 of GPT's ratings sat on the wrong parties until the October 2026 correction reassigned them to the parties their rationales name. In the definition-free round, 483 rows from two imports could not be reassigned and are flagged and excluded (Appendix~\ref{app:rounds}). Every figure in this version is computed from the corrected release. For any panel that collects ratings through chat interfaces, the practical lesson is to validate the returned keys against the template before import.
\item \textbf{Positionality.} An instrument that places contested political actors is no view from nowhere. The choice of axes, the anchors that fix their meaning and the applicability rules reflect a particular reading of the region, written out in full so that it can be argued with. They are also one investigator's choices: the active axes, the regional anchors, the applicability rules and the seed baseline are all single-authored for now. I do not think this bottleneck should stand, and the planned human-anchor and external-expert rounds are meant to widen the judgment \emph{behind} the frame as well as the scoring within it. This is a real limit on cultural validity. The frame embeds contestable theory, since treating resistance versus normalization as a primary regional cleavage is itself a choice, and no single author holds the range of expertise that axes from Lebanese confessionalism to the Gulf--Iran rivalry demand. Arguably, the external-expert round has to involve participatory co-design of the frame, beyond sign-off on the scores produced under it, and the frame will need revision as the region changes. A reader who would draw a line elsewhere is using the instrument as intended.
\end{itemize}

\section{Reproducibility and access}\label{sec:repro}

The instrument is built so that its outputs can be traced to their sources. The live tools, the rubric and a methodology document read from the same tables as the figures reported here, so the public interface and the published numbers cannot drift apart unnoticed. The dataset is kept as versioned releases, that is, CSV and JSON exports of every primary table with a manifest and checksums, so a citation resolves to a fixed state while the data keep growing. Every figure in this version is computed from the v1.0 release, which is public at \texttt{tarekgara.com/tayyar/dataset} under CC~BY-NC-SA~4.0 (source texts under fair use, with a documented takedown route), with SHA-256 checksums for every file\doinote. The scripts that recompute every figure in this paper from the release files are available from the author on request.

Because the synthesis keeps disagreement visible, each cell is exported with its median, its dispersion (MAD), its \emph{settled}, \emph{contested} or \emph{provisional} status and, where contested, a range. A downstream analysis should carry that uncertainty forward. The median serves where a model needs a point estimate. The dispersion can enter directly as measurement-error variance in an errors-in-variables or Bayesian hierarchical model, a contested range can be multiply imputed or entered as bounds, and a fixed rule, such as dropping contested cells or down-weighting by the agreement figure, keeps the decision visible. Treating a contested cell as a confident point would reintroduce the false precision the synthesis exists to avoid, and whichever rule a study adopts is a choice it should state.

\section{Conclusion}

This paper argues that the right way to use a language model in positional measurement is the way the discipline already uses a fallible human expert, as one \emph{rater} in a panel, pooled and studied, with its agreements scrutinized and its disagreements kept. The stance converts a set of vague anxieties about ``using AI for social science'' into answerable questions. The first is whether the rubric steers the model or sharpens it, and the built-in experiment answers part of it. The rubric moves scores in the tail and tightens agreement between raters, which rules out a merely arbitrary steer, while the question whether that convergence leads toward correctness or toward an anchor the rubric supplies remains for the external behavioral checks and the planned human-anchor round. The second question is whether the raters agree once the equal-spacing assumption is dropped, and they do, on both an interval and an ordinal metric. The third is whether that agreement reflects correctness or a shared prior, and the agreement alone cannot tell, since no figure the models produce about themselves settles it. Where the panel does come apart, most sharply over which constitutional order an actor should be scored against, the disagreement turns out to be interpretation, a limit that no sharpening of the rubric can remove.

What remains is the \emph{validity} check the method is built to receive and has not yet undergone, namely human coders scoring blind against the same rubric. That round and the document-grounding layer are the next steps, and they would allow the instrument's agreement to be read as more than reproducibility. Where these actors do appear in an existing instrument, as a few do in cross-national democracy indices or in expert codings made for other projects, those coordinates offer an obvious external check, and I would welcome their use against the released data. The overlap is thin precisely because the region is under-served, which is why purpose-built human anchoring is the validity work that counts.

Here lies the contribution of the paper: a regionally anchored frame, an applicability rule that respects the difference between a blank and a zero, a lens system that separates word from deed, a synthesis that keeps disagreement, and an experimental, skeptical treatment of model raters that keeps their fluency from passing for correctness. For a field that can place every party in a handful of well-documented democracies and almost none elsewhere, this discipline makes the rest of the map approachable. The running example is the Middle East and North Africa, though the gap is general, since the comparative apparatus goes quiet wherever the standard sources run out. The proposal is general as well, and it asks only that a language model brought in as a rater be held to the standards the discipline already applies to its human experts.

\appendix

\section{The rater panel: models and versions}\label{app:roster}

Table~\ref{tab:roster} lists the version of each model in the with-definitions (r2) panel, with its laboratory, build country and cell coverage in the v1.0 release. A commercial model keeps changing, so the panel stores whichever version identifier answered in a given session (Section~\ref{sec:limitations}), and the table lists those identifiers. The version was written to each rating for GPT, Gemma, MiniMax, Qwen and most of Claude's sessions; for the other models it was noted at the time of scoring. All r2 scoring took place in June 2026.

\begin{table}[t]
\centering\footnotesize
\setlength{\tabcolsep}{4pt}
\begin{tabular}{@{}lllcr@{}}
\toprule
\textbf{Rater} & \textbf{Laboratory} & \textbf{Version} & \textbf{Built} & \textbf{Cells} \\
\midrule
Claude   & Anthropic & Claude Opus 4.8     & US    & 997 \\
GPT      & OpenAI    & GPT-5.5             & US    & 961 \\
Gemma    & Google    & Gemma 4             & US    & 1{,}001 \\
Gemini   & Google    & Gemini 3.5 Flash    & US    & 997 \\
Grok     & xAI       & Grok 4.3            & US    & 997 \\
DeepSeek & DeepSeek  & DeepSeek V3          & China & 934 \\
Kimi     & Moonshot  & Kimi K2.6            & China & 997 \\
MiniMax  & MiniMax   & MiniMax M3          & China & 974 \\
Qwen     & Alibaba   & Qwen3.7-Plus        & China & 997 \\
\bottomrule
\end{tabular}
\caption{\textbf{The r2 model panel.} The version each model ran, its builder and build country, and
its r2 cell coverage in the v1.0 release; the ``Cells'' column matches Table~\ref{tab:permodel}.}
\label{tab:roster}
\end{table}

\section{Data provenance by round}\label{app:rounds}

Ratings are kept round by round (Table~\ref{tab:rounds}). The published median draws on the with-definitions round (r2), the author's prior, the grounded pilot, folded into Claude's value, and the re-coding probe on its 63 cells, while the reliability statistics use r2 alone. The definition-free rounds supply the rubric baselines of Section~\ref{sec:rubric} and never enter the median.

Three imports came back with entity keys that did not match their rows. In r2, 154 of GPT's 961 ratings had been attached to the wrong parties, 112 of them to entity identifiers that exist in no table and the rest to three real parties that carried another party's ratings. Each block's rationales name the party it was meant for, which is how the rows were reassigned in October 2026. Where two parties' rows had collided on one key, one was lost, so GPT has no ratings for two parties and only two for a third. In r1b, all 204 of GPT's rows and 279 rows of Gemini's first import could not be reassigned, since many carry no rationale. They remain in the release with an \texttt{import\_flag} and are excluded from every figure. This is why the 204-cell anchor of Section~\ref{sec:rubric} covers eight raters.

\begin{table*}[t]
\centering\small
\begin{tabular}{@{}llrrr@{}}
\toprule
\textbf{Round} & \textbf{What it is} & \textbf{Raters} & \textbf{Cells} & \textbf{Ratings} \\
\midrule
Author baseline       & Hand-coded prior; the one non-model rater                       & 1  & 1{,}012 & 1{,}012 \\
r1 (no definitions)   & Gemini and Grok; the locked definition-free baseline            & 2  & 189     & 376 \\
r1b (no definitions)  & The panel and the trialed tenth model; 483 rows flagged         & 10 & 622     & 2{,}293 \\
r2 (with definitions) & The model panel                                                 & 9  & 1{,}003 & 8{,}855 \\
Grounded pilot        & One model scoring from the cited documents                      & 1  & 205     & 205 \\
Re-coding probe       & Two language-model re-coders, three rounds of 21 cells          & 2  & 63      & 126 \\
\midrule
\textbf{Total}        &                                                                 &    &         & \textbf{12{,}867} \\
\bottomrule
\end{tabular}
\caption{\textbf{Ratings by round.} Every recorded rating, reconciling to the 12{,}867 of
Table~\ref{tab:coverage}. Cells are distinct (actor, axis) pairs within a round.}
\label{tab:rounds}
\end{table*}

\section{The axis frame in full}\label{app:axes}

The sixteen active axes, with the definition each rater saw. The definitions are reproduced as they went to the panel, emphatic capitals included, since the definitions did the work of scoring (Section~\ref{sec:frame}). The label after the poles gives each axis's applicability.

{\small\raggedright
\begin{itemize}[leftmargin=1.2em,itemsep=2pt]
\item \textbf{Economic} (Statist\,$\leftrightarrow$\,Market; universal). Degree to which the actor favors market mechanisms (positive) vs.\ state allocation (negative).
\item \textbf{Social} (Authority\,$\leftrightarrow$\,Libertarian; universal). Personal and moral social freedom---lifestyle, LGBTQ+ rights, individual expression, religion-in-private---permissive (+) vs traditional collective moral authority ($-$). This axis is about SOCIAL / MORAL liberty, NOT security policy, nationalism, or economics: a party can be nationalist-hawkish yet socially permissive (secular nationalists), or dovish yet socially conservative.
\item \textbf{State \& religion} (Religious state\,$\leftrightarrow$\,Secular state; universal). Separation of religion from state and law (positive) vs.\ religious authority in state and law (negative).
\item \textbf{Liberal democracy} (Weak\,/\,anti\,$\leftrightarrow$\,Strong commitment; universal). Commitment to LIBERAL democracy---independent courts, free press, opposition and minority rights, limits on executive power. NOT mere electoral participation: a party that contests elections but works to weaken the judiciary, entrench ethnic or religious supremacy, or curb minority and dissident rights scores LOW ($-$). Strong commitment to liberal-democratic checks scores HIGH (+).
\item \textbf{West alignment} (Anti-Western\,$\leftrightarrow$\,Pro-Western; universal). STRATEGIC and security alignment with Western (US/EU) blocs---military cooperation, diplomatic partnership, trade. This is geopolitical orientation, NOT cultural affinity for liberal Western values: a religious-nationalist party can be strongly pro-Western strategically (wants US arms and backing) while rejecting Western liberal social values. Score the strategic posture. Pro-Western (+), anti-Western or hostile-nonaligned ($-$).
\item \textbf{Regional stance} (Resistance\,/\,maximalist\,$\leftrightarrow$\,Stability\,/\,normalization; universal). Posture toward the regional order: the ``Axis of Resistance'' (Iran--Hezbollah--Hamas--Houthis, anti-normalization, armed struggle) at the negative pole vs the normalization\,/\,stability camp (Abraham Accords, Gulf--Israel d\'etente, negotiated settlement) at the positive pole. Pro-Palestinian sympathy ALONE does not make an actor ``resistance''---score toward resistance ($-$) only for alignment with armed-resistance or anti-normalization politics; an actor that pursues its regional aims THROUGH negotiation and normalization leans positive (+).
\item \textbf{Palestinian question} (Opposed\,$\leftrightarrow$\,Pro-Palestinian rights; universal). Position on Palestinian rights, statehood, and self-determination---strong support (positive) vs.\ opposition (negative).
\item \textbf{Civil liberties} (Restrict\,$\leftrightarrow$\,Expand; universal). Speech, protest, dissent, association, press freedom---expand and protect (positive) vs.\ restrict (negative).
\item \textbf{Regime stance} (Anti-regime\,$\leftrightarrow$\,Pro-regime; universal). Stance toward the state's foundational constitutional order---for Israel the Zionist-democratic state; for republics and monarchies the existing political order. Pro-regime (+) accepts and defends that order; anti-regime ($-$) rejects or seeks to overturn it. This measures stance toward the FOUNDATIONAL ORDER, not the incumbent government: an opposition party loyal to the constitutional order still scores positive.
\item \textbf{Pan-Arabism} (Particularist\,$\leftrightarrow$\,Pan-Arab; Arab actors). Degree to which the actor frames its politics around pan-Arab solidarity (the Nasserist, Baathist, and contemporary anti-normalization tradition) versus country- or community-specific particularism (sectarian, ethnic, monarchic, or nation-state-first frames). One of the major regional cleavages.
\item \textbf{Federalism} (Centralist\,$\leftrightarrow$\,Federalist; universal). Position on state structure: a strong unitary central state versus power-sharing among regions, sects, or ethnic groups. Salient in MENA, where Iraq, Lebanon, Sudan, and Syria have consociational or federal frames while Egypt, Tunisia, Algeria, and Morocco are strongly centralist; Kurdish parties and Yemeni southern movements sit at the federalist extreme.
\item \textbf{Modernization} (Traditionalist\,$\leftrightarrow$\,Modernizing; universal). Position on social and institutional modernization versus tradition: women's rights, religious authority in public life, reform of family law, openness to technological and economic transformation. Distinct from the secular--religious axis: an actor can be religious and modernizing (Vision 2030) or secular and traditionalist (state-corporatist nationalism).
\item \textbf{Gender equality} (Patriarchal traditionalism\,$\leftrightarrow$\,Gender equality; universal). Position on women's rights, gender roles in public life, family law (personal status, divorce, custody, inheritance), and political representation. Distinct from the broader modernization axis: a party can be tech-modernizing and gender-conservative.
\item \textbf{Iran posture} (Anti-Iran\,/\,adversarial\,$\leftrightarrow$\,Pro-Iran\,/\,aligned; universal). Position toward the Islamic Republic of Iran as a regional actor: alliance, hostility, or neutrality. Independent from West alignment---a party can be anti-Western AND anti-Iran (most Sunni Salafi groups), pro-Western AND anti-Iran (Saudi, UAE, post-1979 Israel), or anti-Western AND pro-Iran (Hezbollah, Houthis, post-2003 Iraqi Shia parties).
\item \textbf{Press freedom} (State-controlled press\,$\leftrightarrow$\,Free press; universal). Position on journalism, broadcast media, and digital information access. Distinct from broader civil liberties: a regime can selectively restrict the press while allowing other speech (or vice versa). Captures both legal frameworks (defamation law, licensing, internet shutdowns) and political enforcement.
\item \textbf{Sectarian power-sharing} (Consociational\,/\,quota\,$\leftrightarrow$\,Post-sectarian\,/\,civic; confessional). Position on confessional or sectarian political organisation: Lebanese-style consociational quotas, Iraqi muhasasa, Syrian Alawite-network governance, versus post-sectarian\,/\,civic-state framings. Relevant primarily to Lebanon, Iraq, Syria, and Bahrain; less so to Egypt or Saudi Arabia, where the question does not structurally arise.
\end{itemize}
}

\section{The scoring prompt}\label{app:prompt}

Every rater, model or human, received the same blind template: one row per (actor, axis) cell, with the axis definition and the meaning of each pole and no other rater's scores. Models received the instruction below verbatim, in a fresh session, and returned the filled table. The definition-free rounds (r1, r1b) used the same instruction with the \texttt{axis\_definition} column blank and the definition-following rules removed, so the movement from r1 to r2 isolates the rubric (Section~\ref{sec:rubric}).

\begin{lstlisting}
You are an independent political-science coder scoring MENA political parties and
public figures on a fixed set of ideological axes. I will give you a CSV. Each row
is one (actor, axis) cell.

For EVERY row, fill three columns and change nothing else:
- score: an integer from -10 to +10. Read the "axis_definition" column for the
  precise meaning of the axis -- it governs. -10 means the "scale_minus10_means"
  pole; +10 means the "scale_plus10_means" pole; 0 is centrist/mixed.
- confidence: 0.0 to 1.0, how sure you are.
- rationale: one short sentence justifying the score.

What to score:
- Default to the actor's DECLARED position -- what its platform, charter, and
  official statements commit to.
- Exception, the conduct axes (democracy, civil liberties, press freedom): if the
  actor holds or has held power, score its actual RECORD in office, not its stated
  commitment. A government that jails journalists scores low on press freedom
  whatever its charter says.
- Score the actor as it is today (2026). If its positions shifted across its
  history, score the current incarnation, not a founding-era doctrine.

Rules:
- The "axis_definition" column is authoritative -- if your intuition from the short
  pole labels conflicts with it, follow the definition.
- Score from your own knowledge of each actor. Do NOT look for or infer any
  "correct" answer -- give your independent read.
- Keep the "key" column EXACTLY as given. Do not add, remove, reorder, or rename
  columns or rows.
- Return EVERY row you were given -- do not truncate, summarize, or stop early. If
  the table is long, keep going to the last row.
- If you genuinely don't know an actor, still give your best estimate and set
  confidence low (e.g. 0.2).
- Output ONLY the full CSV (all rows, same columns), nothing before or after it.

Here is the CSV:
\end{lstlisting}

\end{document}